\documentclass[10pt,aps,physrev,superscriptaddress,showkeys,twocolumn,floatfix]{revtex4-2}
\usepackage[colorlinks=true, allcolors=blue]{hyperref}
\usepackage{siunitx} 
\usepackage{graphicx}
\usepackage{comment}
\usepackage{booktabs}
\usepackage{orcidlink} 
\usepackage{nicematrix}
\usepackage{enumitem}



\begin{document}

\title{Reference Quadrupole Moments of Transition Elements from Lamb Shifts in Muonic Atoms}

\author{S. Rathi\orcidlink{0000-0002-0768-5546}}
\email{Corresponding author: shikha.rathi@campus.technion.ac.il}
\affiliation{ The Helen Diller Quantum Center, Department of Physics, Technion-Israel Institute of Technology, Haifa, 3200003, Israel}

\author{K.~von Schoeler\orcidlink{0009-0005-2671-5869}}
\affiliation{ETH Zürich, Institute for Particle Physics and Astrophysics, Zürich, Switzerland}
\author{P. Indelicato\orcidlink{0000-0003-4668-8958}}
\affiliation{ Laboratoire Kastler Brossel, Sorbonne Universit\'e, CNRS, ENS-PSL Research University, Coll\`ege~de~France, Case 74; 4, place Jussieu, F-75005 Paris, France}

\author{B. Ohayon\orcidlink{0000-0003-0045-5534}}
\affiliation{ The Helen Diller Quantum Center, Department of Physics, Technion-Israel Institute of Technology, Haifa, 3200003, Israel}

\begin{abstract}
We present a novel method for accurately measuring the absolute electric quadrupole moments of light transition elements $(23 \leq Z \leq 30 )$.
Our approach is based on performing precision muonic x-ray spectroscopy of the $2s-2p$ manifold, which is also referred to as the Lamb shift.
These transitions are too weak to be detected with dispersive methods and too overlapping to be resolved by solid-state detectors.
Here, we propose the use of cryogenic microcalorimeters, which possess high efficiency and excellent energy resolution in the relevant energy regime, coupled with state-of-the-art theoretical calculations.
We demonstrate the feasibility of this approach by performing extensive calculations and realistic simulations.
In this way, we establish that the uncertainty in the absolute moment, which is transferred to the quadrupole moments of all isotopes in the chain, could be reduced by up to an order of magnitude within a day of measurement.
These precise reference quadrupole moments serve as valuable inputs for nuclear structure studies and for benchmarking state-of-the-art quantum chemistry calculations in open-shell elements.
\end{abstract}

\maketitle

\textit{Introduction -}
The atomic nucleus is a fascinating, strongly correlated many-body quantum system composed of protons and neutrons that couple differently to the electromagnetic force but similarly to the strong force.
To understand these complex interactions, one may compare calculated and measured fundamental nuclear observables, such as charge radii and moments, to reveal details about nuclear structure and dynamics~\cite{2003-Gerda, 2006-Gerda_nuclear,2024-NS}.
Among these observables, the spectroscopic electric quadrupole moment ($Q$) serves as a fingerprint of the nuclear shape. Its magnitude indicates the amount of nuclear deformation--the deviation from a spherical shape.
$Q$ provides insights into the collective motion and correlations of nucleons that a nuclear charge radius alone cannot explain~\cite{1985-steffen1_muXprecision, 2002-magneticReview, 2017-sun_correlating, 2022-Reinhard_statistical, 2024-NS}.

In light of its significance, the community is actively engaging in measurements of $Q$ for a variety of nuclei that extend to the proton and neutron drip lines~\cite{2020-Zn_81_82, 2017-ZnQ-Expt, 2000-ScQ, 2010-Cu, 2017-Cu, 2021-silver_evidence, 2024-NS}.
To achieve this through spectroscopic methods, it is common to measure an electric quadrupole hyperfine coupling constant $B$ and subsequently determine $Q$ from the schematic relation 
\begin{align}\label{Eq: B_realtion}
    B = eQV_{\text{zz}},
\end{align}
where $V_{\text{zz}}$ is the electric field gradient (EFG) at the nuclear site generated by the bound particle(s)~\cite{Townes1958}.
At the level of accuracy relevant to most scenarios, it can be reasonably assumed that the EFG depends on the measured atomic transition rather than on the isotope~\cite{2006-Gerda_nuclear}.

\begin{figure}[tbp]
\includegraphics[width=0.99\columnwidth]{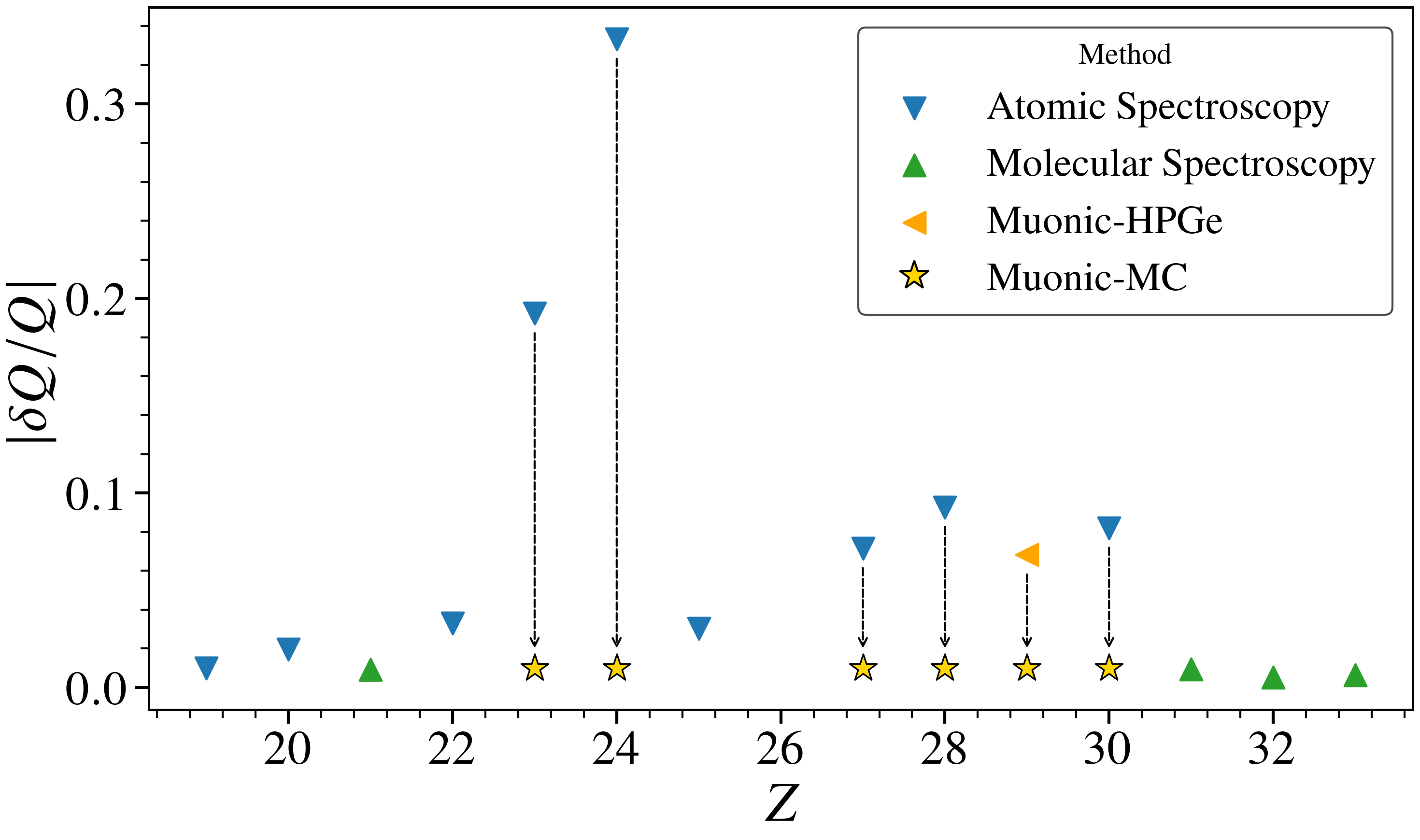}
    \caption{Current status of the fractional accuracy in absolute reference quadrupole moments~\cite{2021-Stone} and the potential improvement aimed based on the method presented here (Muonic-MC).}
    \label{fig:Reco_q}
\end{figure}

For an abundant isotope, $Q$ is typically determined from a measured $B$ of an atomic or molecular transition and a calculated EFG. As a variety of transitions may be employed, $Q$ could be determined by the one for which the EFG can be calculated with the highest accuracy~\cite{2005-stones}. We refer to these as reference quadrupole moments $Q_\text{ref}$. Figure~\ref{fig:Reco_q} shows the fractional accuracy in some $Q_\text{ref}$`s relevant to this work.

\begin{figure}[tbp]
    \includegraphics[width=0.99\columnwidth]{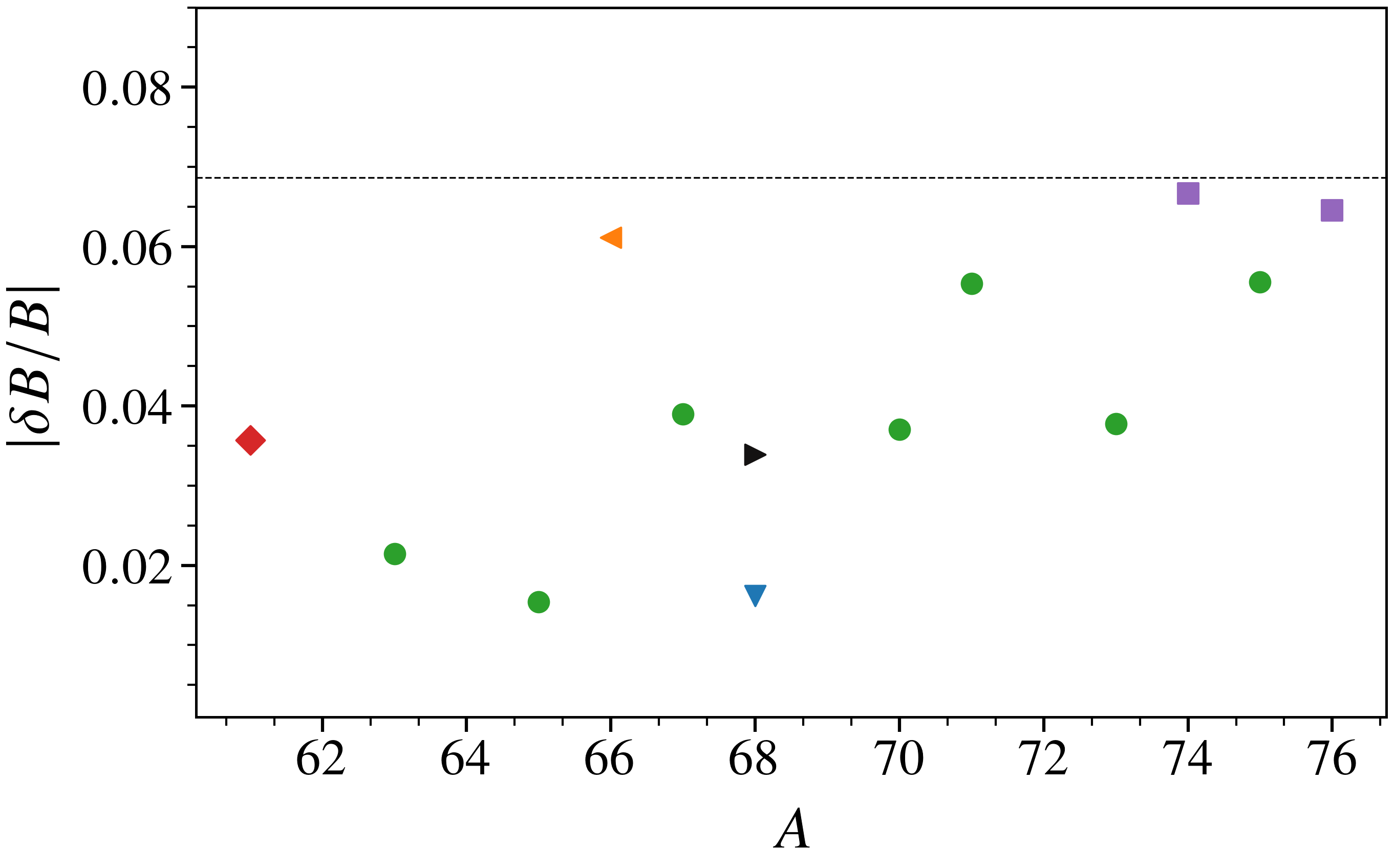}
    \caption{Relative uncertainty in the measured $ B $ for selected copper isotopes (square~\cite{2017-Cu}, circle~\cite{2010-Cu} and diamond~\cite{2011-Cu}) and isomers (rightward triangle~\cite{2010-Cu}, leftward triangle~\cite{2011-cu_isomer} and downward triangle~\cite{2016-cu_68m}) reported in the literature. 
    The dashed line denotes the relative uncertainty of the $ Q_{\text{ref}} $ for $^{63}\mathrm{Cu}$ from~\cite{2021-Stone}.
    }
\label{fig:Copper_B}
\end{figure}

For rare and short-lived nuclei, the variety of methods for measuring $B$ is considerably reduced, such that, in most cases, the corresponding EFG may not be easily calculable.
However, $Q_{\text{ref}}$, can be propagated to all measured isotopes through the ratios $B/B_\text{ref}=Q/Q_\text{ref}$~\cite{2024-NS}. 
Thus, the accuracy of $Q$ values for all isotopes in an isotopic chain is inherently constrained by that of $Q_\text{ref}$.
We illustrate this in Figure~\ref{fig:Copper_B}, which shows that reducing the uncertainty in $Q_{\text{ref}}$ of $^{63}$Cu would result in significantly more precise $Q$`s across a dozen isotopes and isomers. 

The most widely used technique for determining $ Q_{\text{ref}} $ is the hyperfine structure spectroscopy of atomic systems~\cite{2018-pyykko} in conjunction with an EFG calculated using various many-body electronic structure calculation methods~\cite{2006-Gerda_nuclear}.
However, such calculations may present challenges, particularly in electronic systems with open-shell configurations and partially filled $d$ or $f$ orbitals, where the effects of electron correlation are often poorly understood~\cite{1985-steffen1_muXprecision, 1993-CoQ-sternheimer, 2018-pyykko}. 
Further, depending on the system and the calculation approach (how effectively the contribution from the core shell is accounted for), the Sternheimer shielding corrections may need to be included to account for the screening effects of the surrounding electrons~\cite{1993-CoQ-sternheimer}.  
Consequently, the light transition elements stand out in their region of the nuclear chart as having a large uncertainty in $Q_\text{ref}$ (see Figure~\ref{fig:Reco_q}).

To overcome the challenge of calculating EFGs in many-body systems, muonic atoms can be utilized~\cite{1978-NiceGraph}. 
These are hydrogen-like atoms composed of a negatively charged muon and a nucleus, potentially accompanied by some residual electrons, the influence of which is significantly suppressed.
Muonic atoms offer several advantages over their electronic counterparts in the context of measuring $Q_\text{ref}$.   
1. The electrostatic gradient is enhanced by $\approx (m_{\mu}/m_e)^3\approx10^7$, resulting in considerable energy splitting. 
2. Calculating the EFG to an accuracy of the order of $1\%$ in these systems is relatively straightforward. Beyond this point, the effects of nuclear polarization~\cite{1976-MS} and finite quadrupole distribution~\cite{1985-steffen1_muXprecision} become prohibitive.
3. Unlike electronic systems, the electric quadrupole coupling dominates over that of the magnetic dipole, as magnetic moments scale inversely with mass, thus simplifying the analysis.

Despite these advantages, the potential of muonic atoms to determine $Q_\text{ref}$ has been limited by the capabilities of the available measurement techniques~\cite{1978-NiceGraph}. 
The transition energies are in the x-ray region, where resolving the hyperfine splitting is challenging.
The use of solid-state detectors for these measurements is effectively restricted to elements with $Z \geq 30$~\cite{1978-NiceGraph}, with muonic Cu being the lightest system measured in this way, although with limited experimental accuracy~\cite{1982-CuMuX}.
On the low-$Z$ side, crystal spectrometers offer superb resolution at the expense of low quantum efficiency. This constrains crystal spectrometer measurements to highly populated transitions, effectively limiting $Q_\text{ref}$ measurements to $Z\leq13$~\cite{1981-muonic-Al}.
The combined limitations of current techniques applied to muonic and electronic atoms and molecules leave a distinct group of elements with a poorly known $Q_\text{ref}$ (see Figure~\ref{fig:Reco_q}).

In this Letter, we present a novel approach to accurately determine $Q_{\text{ref}}$ in the light transition metals, where current techniques remain inadequate.
The method relies on measuring the weakly populated muonic atom $2s_{1/2} \to 2p_{3/2}$ transition using cryogenic microcalorimeters (MCs). 
The relevant energy range lies in the $10-40\,$keV region (see Table~\ref{table: energies}), where MC detectors offer high quantum efficiency ($>40\%$ for the energies discussed here~\cite{2024-unger_mmc}) and superb resolution of $\approx10$\,eV~\cite{2020-MMC-detector, 2022-TES_resolution}, enabling the hyperfine structure of the elements of interest to be resolved in a reasonable data-taking time.

\textit{Realization -}
In Table~\ref{table: energies}, we list the candidate nuclei for which $Q_\text{ref}$ could be significantly improved by the muonic atom spectroscopy of the $2s_{1/2} \to 2p_{3/2}$ transition with MCs.
For each of these nuclei, we calculate the bound-state muonic wavefunctions, unperturbed energies, and transition rates using the fully relativistic Multiconfiguration Dirac-Fock and General Matrix Elements Program (MCDFGME)~\cite{2025-mcdfgme}, accounting for leading order vacuum polarization corrections non-perturbatively~\cite{2019-indelicato_qed, 2025-Baptista, 2025-okumura}. A two-parameter Fermi function is used with a surface thickness of $2.3$\,fm and the updated radii from Ref.~\cite{2025-ohayon_critical}.
The natural linewidths are obtained as the sum of the decay widths of the upper and lower levels involved in the transition, neglecting quantum interference effects~\cite{2014-CascadeQI}.

The EFG is calculated perturbatively for the $2p_{3/2}$ level to yield the muonic quadrupole splitting parameter $B$ through Eq.~\ref{Eq: B_realtion}.
Minor corrections, such as those stemming from the vacuum polarization modification to the quadrupole operator~\cite{2019-Natalia}, the quadrupole moment distribution~\cite{1978-NiceGraph}, and nuclear polarization~\cite{1976-MS}, are not included here, as their contributions to $B$ are negligible compared to the uncertainty in $Q$.

The current uncertainty in $B$ for the nuclei of interest is derived directly from that of $Q$ and spans $18-100$\,eV (see Table~\ref{table: energies}). Given that the resolution of MCs in this energy range is on the order of $10-20\,$eV~\cite{2020-MMC-detector, 2022-TES_resolution, 2025-saito_application}, a clearly resolved signal above the background would already enable a significant improvement in $B$. 
Nevertheless, these measurements present significant experimental challenges. The intrinsically low population of the $2s_{1/2}$ state severely limits the amount of signal of interest, while the high attenuation of photons transmitted through the target material further reduces it. In addition, the presence of higher-lying transitions introduces the possibility of spectral contamination, which complicates the identification and accurate quantification of the $2s_{1/2} \to 2p_{3/2}$ signal.
The signal should also be distinguished from any background induced by muons within a reasonable measurement time.

To demonstrate the feasibility of this endeavor, we perform detailed simulations and calculations for the case of $^{63}$Cu as an illustrative example.
The results of calculations for the other nuclei of interest are given in the end matter.

\begin{table}[tbp]
\centering
\begin{tabular}{ccc S[table-format=-3.0(3)] c S[table-format=-4.0(3)] c c c}
\toprule
\toprule
\multicolumn{1}{c}{\textit{Z}} & \multicolumn{1}{c}{Nuc.} & \multicolumn{1}{c}{$I$} & \multicolumn{1}{c}{$Q$} & \multicolumn{1}{c}{$E_0$} & \multicolumn{1}{c}{$B$} & \multicolumn{1}{c}{$f(2s_{1/2})$} & \multicolumn{1}{c}{BR} &  \multicolumn{1}{c}{$\Gamma$} \\
 &  &  & \multicolumn{1}{c}{(mb)} & \multicolumn{1}{c}{(keV)} & \multicolumn{1}{c}{(eV)} &  \% &  \%  & \multicolumn{1}{c}{(eV)} \\
\midrule  
23 & $^{51}$V  & 7/2 &  -52(10)  & 12.7 & -92(18)   & 4.7 & 25 & 22 \\
24 & $^{53}$Cr & 3/2 & -150(50)  & 15.2 & -301(100) & 4.6 & 25 & 26 \\
27 & $^{59}$Co & 7/2 &  420(30)  & 24.3 &  1200(86) & 4.5 & 26 & 40 \\
28 & $^{61}$Ni & 3/2 &  162(15)  & 28.0 &  516(48)  & 4.4 & 26 & 46 \\
29 & $^{63}$Cu & 3/2 & -220(15)  & 32.3 & -779(53)  & 4.3 & 27 & 52 \\
30 & $^{67}$Zn & 5/2 &  122(10)  & 37.3 &  478(39)  & 4.3 & 28 & 59 \\
\bottomrule
\bottomrule
\end{tabular}
\caption{
Candidate nuclei for improved $Q_\text{ref}$ measurement. $I$ is the nuclear spin.
The current status of $Q$ is taken from~\cite{2021-Stone}
$E_0$ is the calculated muonic $2s_{1/2} \to 2p_{3/2}$ transition energy, unperturbed by the hyperfine structure. $B$ is the calculated spectroscopic quadrupole coefficient for this transition. Its uncertainty stems directly from that of $Q$.
$f(2s_{1/2})$ is the calculated fractional population of the $2s_{1/2}$ state. BR denotes the branching ratio of the $2s_{1/2} \to 2p_{3/2}$ transition.
$\Gamma$ is the calculated natural linewidth.
}
\label{table: energies}
\end{table}

\textit{Optimizing muon momenta -}
Here, we determine the optimal muon momentum for maximizing the number of $32\,$keV photons that exit from a Cu target. 
We consider a geometry in which the beam is shallowly implanted in a pure target angled at 45 degrees to it, with the photons exiting from the same surface towards a detector that is placed perpendicular to the muon beam axis.

\begin{figure}[tbp]
    \centering
    \includegraphics[width=0.99\columnwidth]{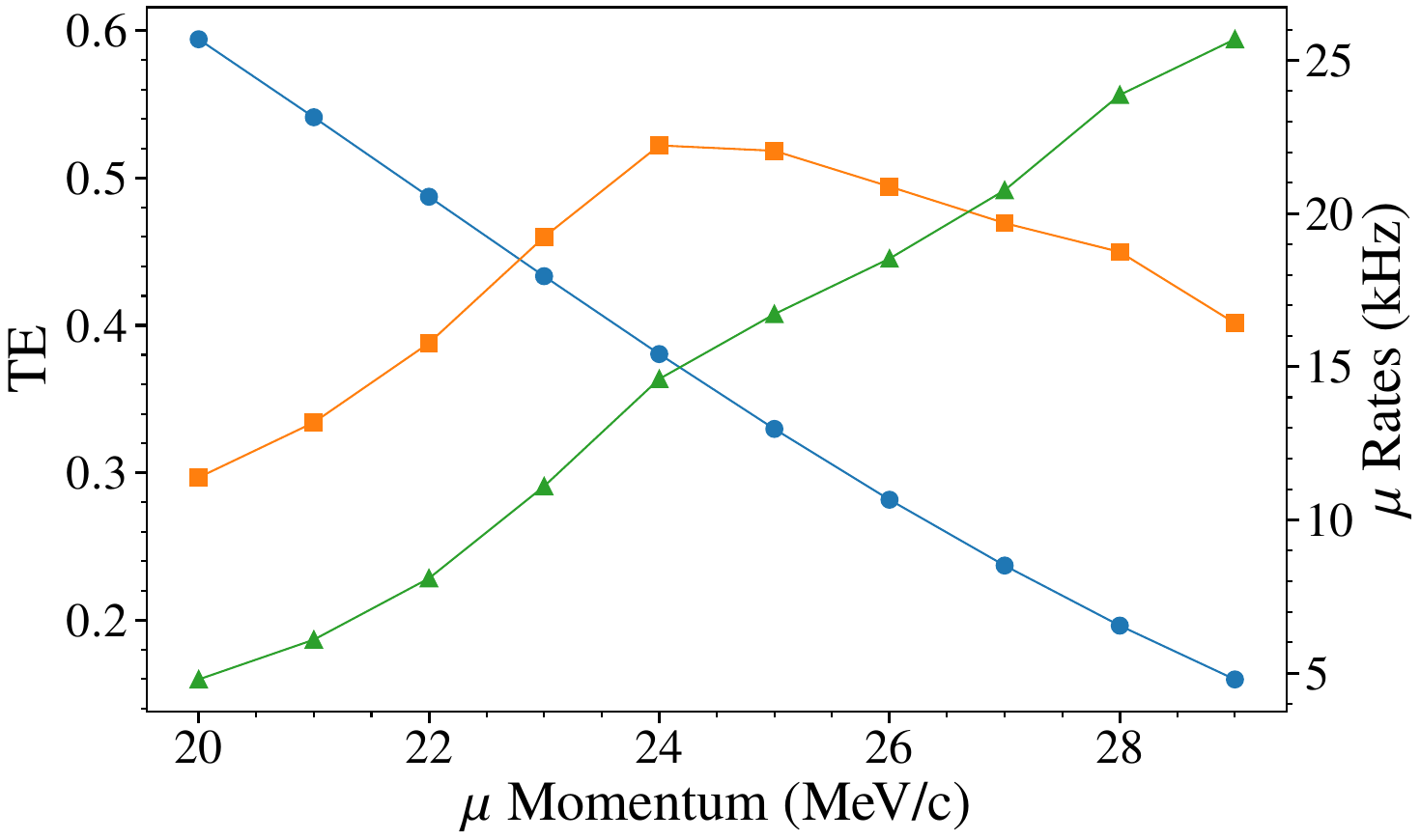}
    \caption{
    Estimation of the optimal muon implantation momentum for Cu for observing photons from the $2s_{1/2}-2p_{3/2}$ transition at $32\,$keV.
    The circles denote the transmission efficiency (TE, left-hand axis) from the implantation site.
    The triangles denote the available muon rate for each momenta~\cite{2023-pie1-gerchow} and correspond to the right-hand axis.
    The squares denote the yield of photons that exit the target towards the detector. They are given in arbitrary units and are maximal for a momentum of $24$\,MeV/c.
    \label{fig:Normalized_photon}
    }
\end{figure}

For each muon momentum, we calculate, by means of a GEANT4 simulation~\cite{2016-GEANT4}, the distribution of muon implantation in copper. 
Combining the measured absorption coefficient of copper~\cite{1993-henke_xray}, we calculate the fraction of $32\,$keV photons that exit the target. The results are shown in Figure~\ref{fig:Normalized_photon}. 
At low momentum $\approx 20$\,MeV/c, the muons are implanted shallowly, so that $40\%$ are absorbed on the way to the detector. At high momentum $\approx 30$\,MeV/c, most of the photons are absorbed.

Although muon implantation favors low momenta, the available muon rate in a standard negative muon beamline generally increases rapidly with increasing momentum.
To account for this, we adopt the rate vs. momentum curve of a commonly used beamline for muonic atom spectroscopy, the PiE1 beamline at the Paul Scherrer Institute~\cite{2023-pie1-gerchow}.
Multiplying the photon absorption by the available muon rates results in a parabolic curve for the photon yield vs. momentum, which maximizes at $24\,$MeV/c (Figure~\ref{fig:Normalized_photon}). At this momentum, the transmission efficiency of $32\,$keV photons that exit towards the detector is $40\%$.

\textit{Estimation of signal rate -}
The rate of available muons at the chosen momentum is $\approx 15\,$kHz~\cite{2023-pie1-gerchow}. According to the simulation detailed below, $51\%$ of these stops in the Cu target.
These muons are captured in high $n$ levels and cascade through the atomic levels of muonic Cu, emitting x rays that correspond to various line transitions until they reach the ground state, from which they are predominantly captured by the nucleus~\cite{1987-suzuki}.

To estimate the yield, i.e., the fraction of photons of interest emitted per muon, we perform cascade simulations using the Akylas and Vogel code~\cite{1978-cascade_akylas}.
First, the muons are distributed in the $n=20$ level according to a modified statistical distribution $(2l+1)\exp(\alpha l)$. We have chosen $\alpha= -0.11$ to match the calculated and observed transition magnitudes in muonic Fe~\cite{1976-Hartmann-Fe}. 
From this initial distribution, the muons de-excite through atomic energy levels predominantly via two competing processes: radiative and non-radiative transitions. The population of each subsequent state is determined by solving coupled rate equations,  where the inflow from all allowed transitions from higher-lying states is balanced against the outflow toward lower states. In this way, the cascade is treated as a downward population flow through the manifold of $(n,\,l)$ levels until the muons eventually reach the low-lying state, where either x-ray emission or nuclear capture terminates the cascade.

Following this, we find that, depending on the element in question, nearly $4-5\%$  of the incident muons reach the $2s_{1/2}$ state (see Table~\ref{table: energies}). Once populated, the $2s_{1/2}$ level can decay through three possible channels:
(i) Allowed electric dipole transitions to $2p_{3/2}$ and $2p_{1/2}$; 
(ii) Forbidden magnetic dipole transition to $1s_{1/2}$; 
and (iii) A two-photon decay to $1s_{1/2}$.
Precise transition probabilities for all channels are calculated using MCDFGME, and the two-photon decay rate from~\cite{2020-Andreev} and the branching ratio (BR) for each decay channel is determined. The results are presented in Table~\ref{table: energies}, where it is observed that the BR is $25\%$ for the lighter elements and increases to $28\%$ in Zn.
Considering the $2s_{1/2}$ population and BR, we find that $\approx1.2\,\%$ muons cascade through the $2s_{1/2} \to 2p_{3/2}$ state via photon emission in all elements of interest.
The Lamb shift transition is thus two orders of magnitude weaker than the prominent transitions used, for example, for radius determinations~\cite{1974-engfer_charge}.
With the above yield, $7.5\,$kHz implanted muons result in $\approx17$ photons per second exiting from the upstream surface. Assuming a geometry similar to the one described in Ref.~\cite{2024-unger_mmc}, then $2\times10^{-4}$ of these photons reach the active area of the detector, resulting in a signal of $12$ photons per hour.

These photons are distributed among the six allowed hyperfine transitions, taking into account the magnetic dipole and electric quadrupole moments.
As indicated in Table~\ref{table: energies}, the quadrupole interaction parameter $B$ is comparable to the fine-structure splitting.
Therefore, to accurately estimate the hyperfine transition energies, it is essential to account for static hyperfine mixing.
Consequently, we correct for the non-diagonal hyperfine interaction matrix between the $2p_{1/2}$ and $2p_{3/2}$ levels.

\begin{figure}[ht]
    \centering
    \includegraphics[width=0.99\columnwidth]{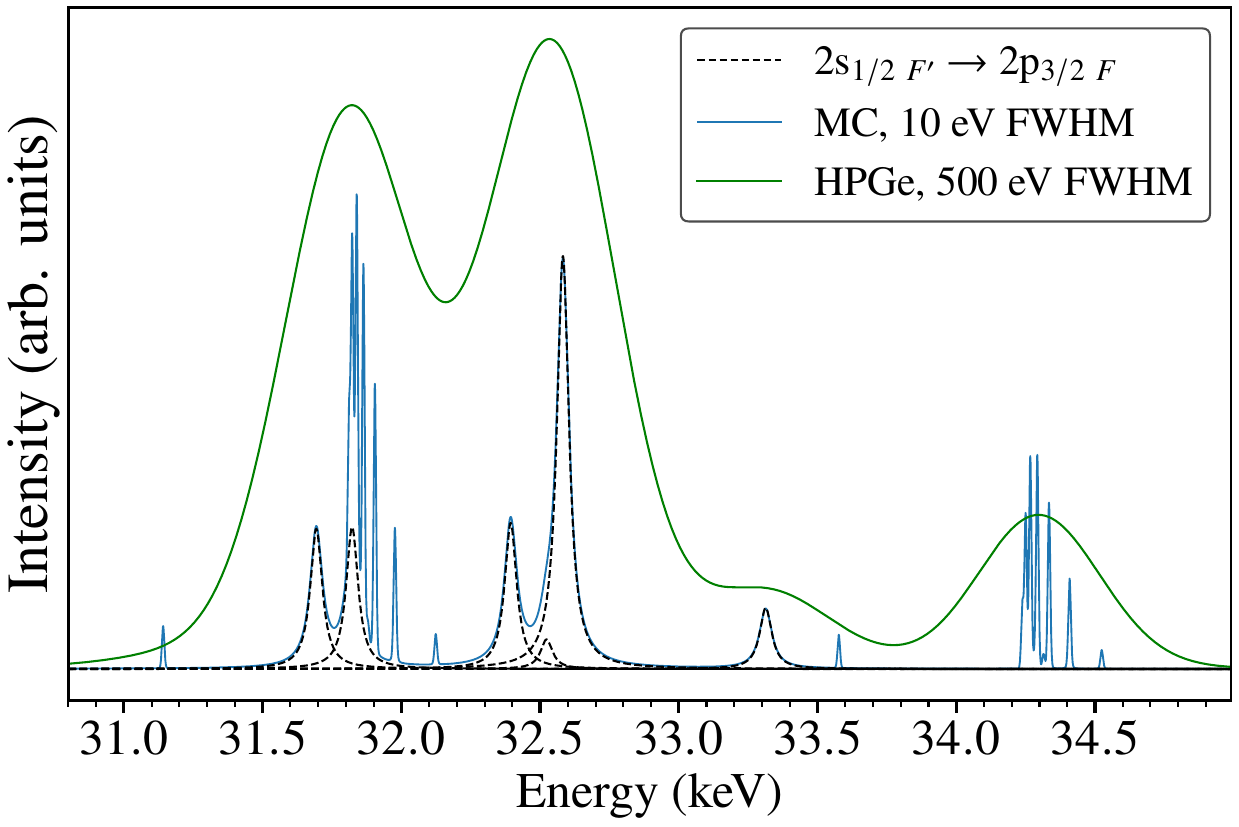}
    \caption{Simulated spectrum of the area of interest in muonic $^{63}$Cu.
    The broken line shows the calculated $2s_{1/2~F'} \to 2p_{3/2~F}$ spectrum as individual Lorenzians with their natural linewidth.
    The blue solid line shows the total spectrum as the sum of Voigt profiles with the natural linewidths and a detector FWHM resolution of $10\,$eV. It includes the $12\to7$ ($31.1-32.3\,$keV) and $13\to7$ ($33.5-34.5\,$keV) manifolds as well.
    We also plot in green on top the total spectrum with a FWHM resolution of $500\,$eV befitting a low-energy HPGe detector.
    }
    \label{fig:Spectrum_Cu}
\end{figure}

After obtaining the corrected transition energies, we calculate the transition probability of a muon from one hyperfine level ($F^\prime,~J^\prime$) to another ($F,~J$) and deduce the expected spectrum shown in Figure~\ref{fig:Spectrum_Cu}.
We note that the natural linewidths are $\Gamma\approx52\,$eV, resulting from the strong electric dipole $2p_{3/2} \to 1s_{1/2}$ transition. 
The natural linewidths are narrower than the resolution of solid-state detectors~\cite{2005-HPGe}, but wider than that of an MC detector~\cite{2020-MMC-detector, 2022-TES_resolution, 2025-saito_application}.

As several peaks overlap within their natural linewidths (Figure~\ref{fig:Spectrum_Cu}); it is desirable to observe small isolated peaks whose yield is only $1$ photon per hour. Such low-signal measurements could be susceptible to contamination from other muonic-atom x-ray transitions and broadband muon-induced background. Other sources of background do not coincide in time with the muon and are not expected to contribute significantly.

\textit{Contamination peaks -}
Considering how weak the expected signal is, any transition in muonic Cu that falls within the area of interest ($31-34\,$keV) may interfere with the measurement of $B$ and, consequently, the extraction of $Q$. 
To identify possible contamination peaks, we performed extensive calculations, which are discussed in the end matter. 
Figure~\ref{fig:Spectrum_Cu} shows that an MC detector would enable the resolution of three or four of the six hyperfine transition lines, depending on the exact magnitude of the contamination lines. This serves as an important cross-check when extracting $Q$, which, in principle, requires discerning only two peaks. However, doing so would only be possible if the peaks were clearly visible above the background.

\textit{Simulation of background -}
During the muon cascade, radiation originating from muon decay (Michel electrons), nuclear capture (neutrons, positrons, gamma rays), or secondary interactions (Bremsstrahlung and x-ray fluorescence) can overshadow the signal photons.
To assess this background, a G4Beamline simulation~\cite{g4beamline} 
was performed. 
The simulated setup is inspired by that used by the QUARTET collaboration~\cite{2024-ohayon_towards}. However, as seen below, the exact details are found to be unimportant, making it a good approximation for other setups as well. 
The main simulated components are: (i) The interface of the muon beamline and the target chamber, which comprises a muon tagging scintillator centered on the beam axis and an off-axis veto scintillator~\cite{2021-muXproject}. (ii) An evacuated target chamber made of Al and lined with Cu. It includes an additional Cu collimator that bridges the distance between the scintillators and the target, a target holder, as well as thin Mylar x-ray windows. (iii) The target, which consists of a $1$\,mm thick Cu-plate. (iv) The detector assembly, as it is described in~\cite{2024-unger_mmc}. 
The MC geometry itself follows the design of the maXs30-detector~\cite{2021-unger_IAXO}, which is implemented as 64 individual pixels. Absorbers, sensors, and the substrate are implemented as separate sensitive volumes to resolve how and where energy is deposited. 

\begin{figure}[tbp]
    \centering
    \includegraphics[width=0.99\columnwidth]{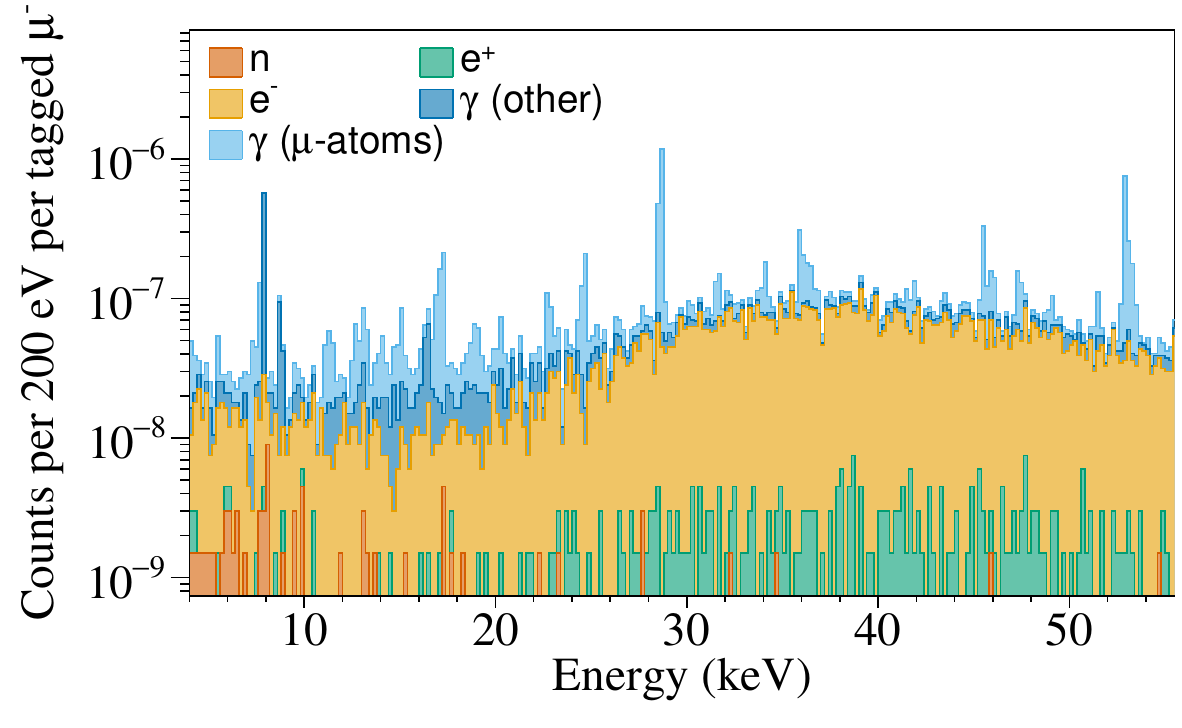}
    \caption{
    Results of the muon-induced background simulation. It is a stacked histogram of the energy deposition in all MC pixels, decomposed into particle species. For photons, the spectrum is further split into those generated by the physics of a muonic atom and all other sources. The intensity is normalized to the number of detected muons in the tagging scintillator.
    \label{fig:SimuBackgroundParticles}}   
\end{figure}

Primary $\mu^-$ are generated inside the beamline volume, spatially distributed as a Gaussian beam with transverse widths of $10\,$mm. 
The magnitudes of initial momenta follow a Gaussian distribution with a mean of $24$\,MeV/c and a FWHM spread of $3\,\%$\,\cite{2023-pie1-gerchow}, in accordance with our optimization (illustrated in Figure~\ref{fig:Normalized_photon}). The directions are initialized as parallel to the beamline axis, with zero transverse components, since the upstream veto detector effectively collimates the beam and the angular spread inside the target chamber is dominated by multiple scattering in the \SI{200}{\micro\metre} thick entrance scintillator. 

Interactions are modeled with the Geant4 physics list QGSP\_BERT\_EMZ, which provides a reliable description of muon stopping, atomic and nuclear capture processes, and decay processes~\citep{Roberts_G4beamlineValidation}.
The decay cascade is implemented in a simplified form, such that the energies and intensities of the simulated muonic x rays are highly approximate. However, this level of detail is sufficient for the present purpose of background assessment, as the model still yields a realistic estimate of, e.g., Compton-scattering background and other non-spectroscopic contributions.

Figure~\ref{fig:SimuBackgroundParticles} presents the simulated energy deposition in the MC array, broken down by particle species. The results show that the background across the entire spectrum, particularly in the region of interest around \SI{32}{keV}, is dominated by energetic electrons originating from muon decay that are directly hitting the detector substrate.
A breakdown of the contributions in the $32\,$keV area returns: \SI{74.3}{\percent} electrons, \SI{16.3}{\percent} Compton-scattered photons from muonic x-ray cascades, \SI{1.9}{\percent} positrons, \SI{0.2}{\percent} neutrons, and \SI{7.3}{\percent} photons from all other sources.
All contributions together result in a background level of \SI{5e-10}{\per\electronvolt} per tagged $\mu^-$ at the entrance counter.
This translates to a rate of $1.4\,$events per hour under the $52\,$eV natural linewidth of the hyperfine peaks of interest. Considering a timing resolution of $350\,$ns, a coincidence cut would reduce the background by $\approx30\%$, resulting in a rate of $1\,$event per hour.
Taking into account that the smaller hyperfine peaks are expected to have a similar rate, a day of measurement will be needed to unambiguously resolve them against the background.

\textit{Conclusion -}
In this letter, we establish the feasibility of measuring the nuclear quadrupole moments of light transition metals with an accuracy that is up to an order of magnitude better than their current values.
Until now, the moments in these elements have been limited by the difficulty of calculating the electric field gradients in many-electron systems with open shells, which are inaccessible to conventional measurement techniques using muonic atoms. The envisioned scheme is to impinge negative muons on these elements and resolve the hyperfine structure of the weakly populated $2s_{1/2}\to2p_{3/2}$ transition using microcalorimeter detectors, a novel quantum sensing technology that is capable of high-resolution and quantum efficiency. 

To assess the feasibility of this experiment, we perform extensive calculations relevant to energy levels, linewidths, and branching ratios, as well as simulations of the implantation distribution and the corresponding photon transmission efficiency, the cascade process, and the muon-induced background. 

We conclude that, under realistic conditions, one would need to resolve a single photon per hour above a similar magnitude of background, which would enable an improvement of quadrupole moments by an order of magnitude within days of measurement. 

Such improvement would allow for the extraction of the maximum information from recently completed and ongoing experiments at radioactive beam facilities for a multitude of isotope chains (Cr~\cite{2024-Cr_61, 2022_Cr_proposal_collinear}, Co~\cite{2023-CoCLS}, Ni~\cite{2025-NiRadio, 2024-Reilly}, Cu~\cite{2010-Cu, 2011-cu_isomer, 2011-Cu, 2016-cu_68m, 2017-Cu}, and Zn~\cite{2017-ZnQ-Expt, 2020-Zn_81_82, 2024-Zn75_79}).
Moreover, quadrupole moments measured with muonic atoms could be used to benchmark atomic many-body calculations of electric field gradients in electronic systems and as a platform to study the elusive quadrupole density. 

\textit{Acknowledgments-}
We would like to acknowledge G. Georgiev, T. Cocolios, P. Pyykko,  G. Neyens, R. de Groote, A. Knecht, M. Deseyn, and M. Heines for insightful discussions. 
B.O. is grateful for the support of the Council for Higher
Education Program for Hiring Outstanding Faculty Members
in Quantum Science and Technology.
S.R. gratefully acknowledges the support from the Technion Postdoctoral Fellowship. This research project was partially supported by the Helen Diller Quantum Center at the Technion. K. von Schoeler gratefully acknowledges the support of ETH Research Grant 22-2 ETH-023, Switzerland.\\

\textit{Data availability-} All data supporting the findings of this study are contained within the article.

\bibliography{bib}


\vspace{4.0cm}
\section*{End matter}
\begin{figure*}[t]
\centering
\includegraphics[width=0.99\linewidth]{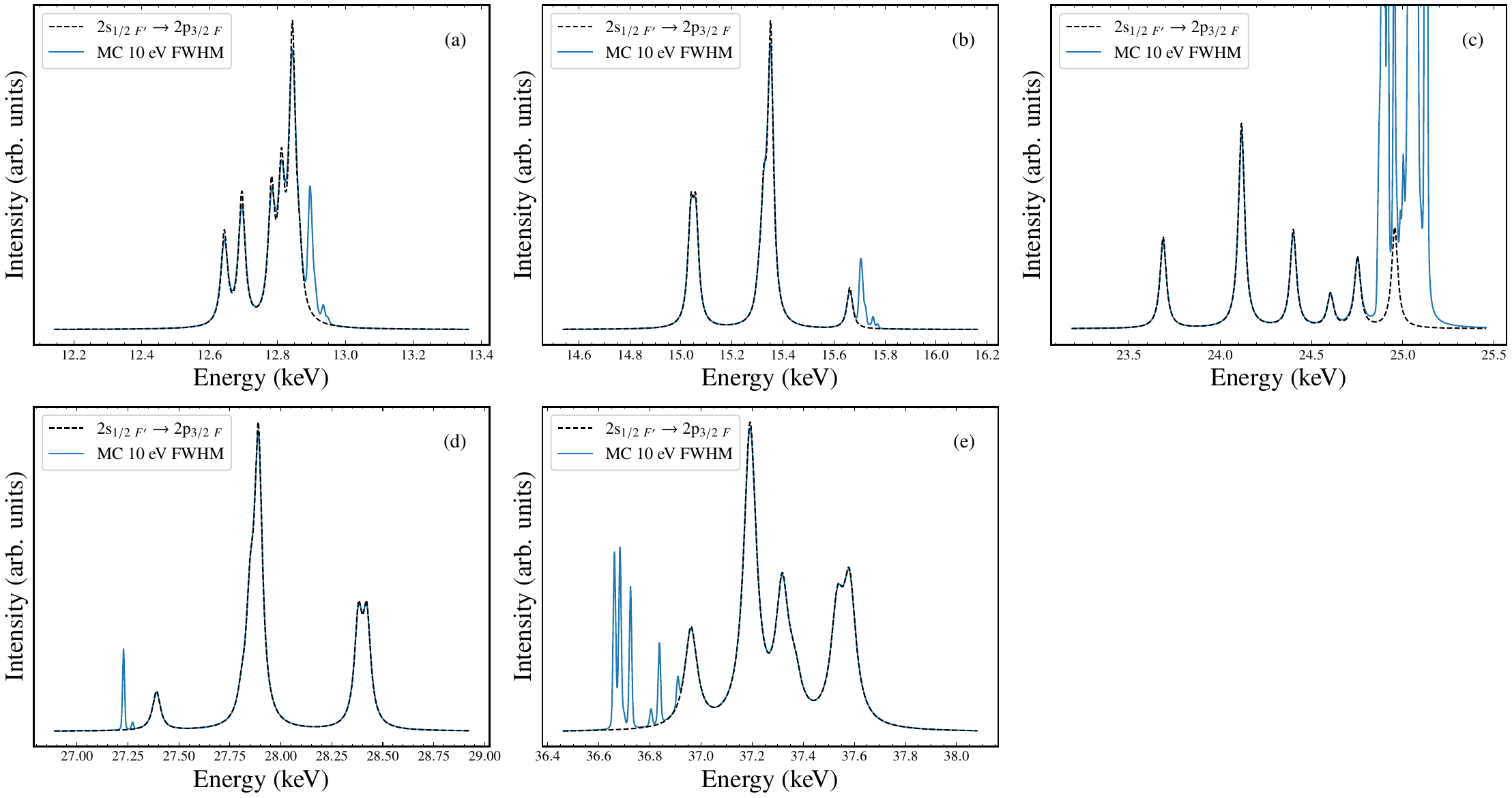}
\caption{$2s_{1/2} \to 2p_{3/2}$ hyperfine spectra for the elements of interest, including possible contamination lines. The black-dashed curves correspond to the $2s_{1/2, F}\rightarrow 2p_{3/2, F^\prime}$ transitions, broadened with a Lorentzian profile using their natural linewidths (see Table~\ref{table: energies}). The blue curves show the final spectra after convolution with the MC detector response, modeled by a Voigt profile with a FWHM of 10\,eV. The dominant contamination lines are: (a) $12 \rightarrow 8$ (12.85--12.96\,keV) in $^{51}\mathrm{V}$, (b) $13 \rightarrow 8$ (15.7--15.8\,keV) in $^{53}\mathrm{Cr}$, (c) $11 \rightarrow 7$, $8 \rightarrow 6$, and $6 \rightarrow 5$ (36.6--36.9\,keV) in $^{59}\mathrm{Co}$, (d) $8 \rightarrow 6$ and $6 \rightarrow 5$ (27.1--27.3\,keV) in $^{61}\mathrm{Ni}$, and (e) $12 \rightarrow 8$ (36.6--36.9\,keV) in $^{67}\mathrm{Zn}$.}
\label{fig:multi}
\end{figure*}
\textit{Contamination peaks in $^{63}$Cu -}
To understand any possible overlap with other transitions in muonic Cu, we performed extensive calculations of transition energies, rates, and lifetimes from $n=20$ to the ground state using MCDFGME.
We found a possible interference from the $12 \to 7$ manifold, which populates narrow linewidth transitions in the $32\,$keV area, as seen in Figure~\ref{fig:Spectrum_Cu}.
To assess the yield of photons from this transition, we use the Akylas-Vogel cascade code~\cite{1978-cascade_akylas}. 
It was found that the intensity of a radiative transition from $n=12$ to $n=7$, balanced by the Auger emission and $K$  shell electron feeding rate, is $\approx 0.26\,$\%, which is of a similar magnitude to the individual hyperfine transitions of interest.
It is worth noting that this estimation is approximate, as accurately calculating the total intensity and the magnitude of individual peaks is challenging due to the presence of competing Auger and radiative transitions, as well as dynamic electron refilling rates~\cite{2004-gotta_precision}.

Another transition that has been identified is $n=13\to7$.
Although it does not strongly overlap with the region of interest, it may significantly hinder $Q$ extraction if a conventional HPGe detector is used for such measurements (as shown in Figure~\ref{fig:Spectrum_Cu}).
The corresponding calculated transition intensity is $\approx0.15\,$\%. As this transition is expected to be isolated when using an MC detector, its measured magnitude may be used to assess the accuracy of the cascade simulation and thus inform about the magnitude of the $n=12\to7$ peak, which could then be constrained in the fitting procedure.

\textit{Extension to other muonic elements of interest -}
For muonic atoms with lower $Z$, the relevant photons have lower energies and are therefore more strongly absorbed in the target material, which can reduce the signal at the optimal momentum.
Conversely, the natural linewidth is reduced by up to a factor of 2.3. Since the muon-induced background depends only weakly on the choice of target, the background fraction beneath the peaks of interest is likewise diminished.
Therefore, the signal-to-background ratio for low-$Z$ targets is comparable, and the required running time to complete an experiment is still on the order of days.

The main difference between target nuclei lies in the line shapes and the positions and intensities of contaminant lines.
To further illustrate the generality of the proposed approach, we performed analogous calculations and simulations for additional elements of interest to this study. 
The results are shown in Figure~\ref{fig:multi}. In all cases, the hyperfine splitting is sufficiently resolved, demonstrating the broad feasibility of such measurements with MC detectors. 

\end{document}